\documentclass[twocolumn,aps,prl,10pt]{revtex4-2}
\usepackage{bm}
\usepackage{amsfonts}
\usepackage{amssymb}
\usepackage{amsmath}
\usepackage{graphicx}
\begin{document}

\title{Classical-quantum correspondence for particles in the Penning trap}
\author{Iwo Bialynicki-Birula}\email{birula@cft.edu.pl}
\affiliation{Center for Theoretical Physics, Polish Academy of Sciences\\
Aleja Lotnik\'ow 32/46, 02-668 Warsaw, Poland}
\author{Zofia Bialynicka-Birula}
\affiliation{Institute of Physics, Polish Academy of Sciences\\
Aleja Lotnik\'ow 32/46, 02-668 Warsaw, Poland}
\date{\today}

\begin{abstract}
We derive new solutions of the Schr\"odinger equation which describe the motion of particles in the Penning trap. These solutions are direct counterparts of classical orbits. They are obtained by injection of classical trajectories into the wave functions of stationary solutions.
\end{abstract}

\maketitle

\section{Introduction}

Penning trap is undoubtedly ``a versatile tool for precise experiments in fundamental physics'' \cite{bnw}. All these experiments involve quantum particles, electrons, ions, etc., but their theoretical description \cite{bg,dbbs,mgw,vog,fmsg} often uses the notion of classical trajectories. The complete set of wave functions of particles in the Penning trap is well known (cf. (\ref{eigf})) but these functions are not related to classical trajectories. Our new solutions of the Schr\"odinger equation in the Penning trap describe quantum wave packets whose centers move along classical trajectories.

A detailed theoretical analysis of the {\em classical dynamics} in the Penning trap is given in \cite{mk}. In general, the classical trajectories are not periodic, with the exception of the very special case of commensurate three frequencies. In contrast, in quantum mechanics the solutions of the Schr\"odinger equation obtained by the separation of variables describe stationary states. Stationary states are always periodic with the frequency determined by the energy. These states do not resemble the classical trajectories shown in Fig.~1, because they do not exhibit any motion; expectation values of all observables are time-independent.

Our solutions of the Schr\"odinger equation for a particle in the Penning trap will be built on stationary states, but they will have particle trajectories embedded in them so that the centers of the wave packets move exactly along the classical trajectories. To construct these solutions we use the ICT method (injection of classical trajectories) described in our recent paper \cite{bb0}. The origins of this method can be traced to earlier papers \cite{gpv,bb}.

We begin with a brief description of the ICT method. Next, we describe the motion of a charged particle in the Penning trap according to classical mechanics and according to quantum mechanics. Finally, we generate new solutions of the Schr\"odinger equation with the use of the ICT method. These solutions unify the classical and quantum descriptions: classical trajectories are embedded in the wave functions obeying the Schr\"odinger equation. From every stationary solution of the Schr\"odinger equation we can build a plethora of new solutions by choosing an arbitrary classical trajectory from a six-parameter family. This procedure results in the perfect realization of the Ehrenfest theorem. The centers of the wave packets follow exactly the classical trajectory and the shape of the wave packet does not change in time.

\section{The ICT method}

The ICT method works for every system with a Hamiltonian which is a quadratic function of the canonical variables,
\begin{align}\label{ham0}
\hat{H}=\frac{1}{2}\hat{p}_iA^{ij}\hat{p}_j+\frac{1}{2}\hat{x}^iB_{ij}\hat{x}^j
+\hat{p}_iC^i_{\;j}\hat{x}^j.
\end{align}
The theorem proved in \cite{bb0} states:
From every solution $\psi$ of the Schr\"odinger equation with the Hamiltonian (\ref{ham0}), $$i\hbar\partial_t\psi(x_1,\dots,x_n,t)=\hat{H}\psi(x_1,\dots,x_n,t),$$ and for every solution $\{x^i(t),p_k(t)\}$ of the classical equations of motion with the same Hamiltonian we can generate a new solution of the Schr\"odinger equation $\psi_{ICT}(x_1,\dots,x_n,t)$ defined by the action of a unitary operator $\hat{U}$ on $\psi$,
\begin{widetext}
\begin{align}\label{ict}
\psi_{ICT}(x_1,\dots,x_n,t)=\hat{U}\psi(x_1,\dots,x_n,t)=\exp\left(-\frac{i}{2\hbar}x^i(t)p_i(t)
\right)\exp\left(\frac{i}{\hbar}x^ip_i(t)\right)
\psi(x_1-x_1(t),\dots,x_n-x_n(t),t).
\end{align}
\end{widetext}
The arguments of the original wave function $\psi$ in this formula are shifted by the classical trajectory. In a more vivid language, the classical trajectory is {\em injected} into the wave function. Since the first two terms in (\ref{ict}) modify the phase, the only change in the probability density $\rho_{ICT}=|\psi_{ICT}|^2$ is the shift of the arguments,
\begin{align}\label{rho}
\rho_{ICT}(x_1,\dots,x_n,t)=\rho(x_1-x_1(t),\dots,x_n-x_n(t)).
\end{align}
This property follows from the fact that for all stationary states the probability density $\rho$ and the probability current $\bm{j}$ do not depend on time. The change of the probability current is,
\begin{align}\label{cur}
&\bm{j}_{ICT}(x_1,\dots,x_n,t)=\bm{j}(x_1-x_1(t),\dots,x_n-x_n(t))
\nonumber\\
&+\frac{\bm{p}(t)}{m}\rho(x_1-x_1(t),\dots,x_n-x_n(t)).
\end{align}
The second term guarantees the validity of continuity equation.

The ICT method provides a unification of the quantum description of states in terms of the wave functions with the classical description in terms of trajectories.
This method is particularly well suited for the motion in the Penning trap because it unifies the classical and quantum descriptions which at the first sight have nothing in common.
\begin{figure}[t]
\begin{center}
\includegraphics[width=0.45\textwidth,
height=0.25\textheight]{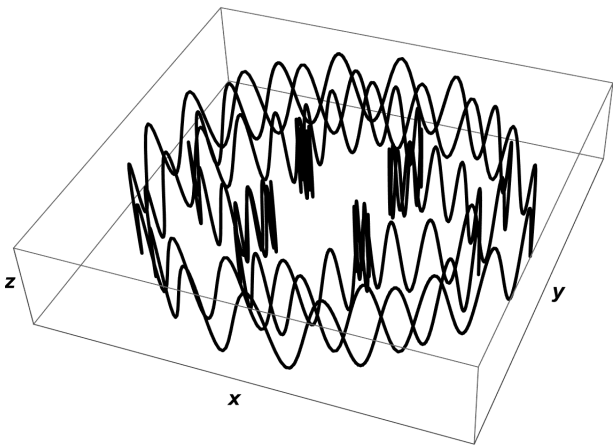}
\caption{Typical trajectory of a charged particle in the Penning trap.} \label{fig1}
\end{center}
\end{figure}
The ICT method also works in the momentum representation; it just involves the interchange of position and momentum variables.

\section{Classical trajectory embedded\\in the quantum mechanical\\wave function}

The probability density $\rho_{ICT}$ defined in (\ref{rho}) depends on time only through classical trajectories. This makes it possible to find the motion of the center of the wave packet, without doing any detailed calculations. The average values of $\bm{r}$ can be easily calculated for any wave function with an injected classical trajectory.
\begin{align}\label{av}
\langle \bm{r}\rangle &=\int\!d^3r\,\bm{r}\rho_{ICT}(\bm{r},t)\nonumber\\
&=\int\!d^3r\,(\bm{r}+\bm{r}(t))\rho(\bm{r},t)=\bm{r}(t).
\end{align}
The integral of $\bm{r}\rho$ vanishes because after the shift $\rho$ becomes an even function of $\bm{r}$. Therefore there are no quantum corrections to the motion of the center of the wave packet; the center follows the injected classical trajectory. In addition, the shape of the wave packet remains unchanged during the time evolution because all central moments of the probability distribution $\rho_{ICT}$ do not change in time,
\begin{align}\label{mom}
\int\!d^3r&(x^i-x^i(t))(x^j-x^j(t))\dots(x^k-x^k(t))
\rho_{ICT}(\bm{r},t)\nonumber\\
&=\int\!d^3r\,x^ix^j\dots x^k\rho(x,y,z)=const.
\end{align}
Hence, the wave packets move along classical trajectories without any change in their shape, as a perfect realization of the Ehrenfest theorem \cite{pe}.

\section{Classical trajectories\\in the Penning trap}

Since the ICT method uses canonical variables, we will use the canonical formulation. The Hamiltonian for a charged particle in the Penning trap is (the $z$ axis is chosen in the direction of the magnetic field $\bm{B}$),
\begin{align}\label{ham}
H&=\frac{1}{2m}\left(\bm{p}-\frac{e}{2}\bm{B}
\times{\bm{r}}\right)^2+\frac{D}{2}(2z^2-x^2-y^2),
\end{align}
where $D$ measures the strength of the electric field. This Hamiltonian describes two uncoupled subsystems described by the sum $H=H_\perp+H_z$,
\begin{align}\label{ham1}
H_\perp=\frac{p_\perp^2}{2m}&+\frac{m\omega_\perp^2x_\perp^2}{2}
-\frac{\omega_c}{2}(xp_y-yp_x),\\
H_z&=\frac{p_z^2}{2m}+\frac{m\omega_z^2z^2}{2}.
\end{align}
where $x_\perp^2=x^2+y^2,\;p_\perp^2=p_x^2+p_y^2$ and
\begin{align}\label{omega}
\omega_\perp^2=\frac{e^2B^2}{4m^2}-\frac{D}{m},\quad
\omega_c=\frac{eB}{m},
\quad\omega_z^2=\frac{2D}{m}.
\end{align}
To keep the particle in the trap the magnetic field and the electric field must satisfy the inequality $e^2B^2>4mD$. Since the Hamiltonian in the Penning trap is quadratic, everything that we described in the previous Sections 

The canonical equations of motion are,
\begin{subequations}\label{ceq}
\begin{align}
\frac{dx(t)}{dt}&=\frac{p_x(t)}{m}+\frac{\omega_c}{2} y(t),\\
\frac{dy(t)}{dt}&=\frac{p_y(t)}{m}-\frac{\omega_c}{2} x(t),\\
\frac{dp_x(t)}{dt}&=-m\,\omega_\perp^2x(t)
+\frac{\omega_c}{2} p_y(t),\\
\frac{dp_y(t)}{dt}&=-m\,\omega_\perp^2y(t)
-\frac{\omega_c}{2}p_x(t),\\
\frac{dz(t)}{dt}&=\frac{p_z(t)}{m},\\
\frac{dp_z(t)}{dt}&=-m\,\omega_z^2z(t).
\end{align}
\end{subequations}
The general solution of these equations is,
\begin{widetext}
\begin{subequations}\label{sol}
\begin{align}
x(t)&=\cos\left(\frac{\omega_c}{2}t\right)\left(x_0\cos(\omega_\perp t)+\frac{p_{x0}}{m\omega_\perp}\sin(\omega_\perp t)\right)+\sin\left(\frac{\omega_c}{2}t\right)\left(y_0\cos(\omega_\perp t)+\frac{p_{y0}}{m\omega_\perp}\sin(\omega_\perp t)\right),\\
y(t)&=\cos\left(\frac{\omega_c}{2}t\right)\left(y_0\cos(\omega_\perp t)+\frac{p_{y0}}{m\omega_\perp}\sin(\omega_\perp t)\right)-\sin\left(\frac{\omega_c}{2}t\right)\left(x_0\cos(\omega_\perp t)+\frac{p_{x0}}{m\omega_\perp}\sin(\omega_\perp t)\right),\\
p_x(t)&=\cos\left(\frac{\omega_c}{2}t\right)\left(p_{x0}\cos(\omega_\perp t)-m\omega_\perp x_0\sin(\omega_\perp t)\right)+\sin\left(\frac{\omega_c}{2}t\right)\left(p_{y0}\cos(\omega_\perp t)-m\omega_\perp y_0\sin(\omega_\perp t\right),\\
p_y(t)&=\cos\left(\frac{\omega_c}{2}t\right)\left(p_{y0}\cos(\omega_\perp t)-m\omega_\perp y_0\sin(\omega_\perp t)\right)-\sin\left(\frac{\omega_c}{2}t\right)\left(p_{x0}\cos(\omega_\perp t)-m\omega_\perp x_0\sin(\omega_\perp t)\right),\\
z(t)&=z_0\cos(\omega_z t)+\frac{p_{z0}\sin(\omega_z t)}{m\omega_z},\\
p_z(t)&=p_{z0}\cos(\omega_z t)-m\omega_z z_0\sin(\omega_z t),
\end{align}
\end{subequations}
\end{widetext}
where the parameters with the subscript 0 denote the initial values of positions and momenta. The trajectory in Fig.1 depicts a typical example of the classical motion.

\section{Solutions of the Schr\"odinger equation in the Penning trap}

The Schr\"odinger equation for a particle in the Penning trap in our notation reads,
\begin{widetext}
\begin{align}\label{se}
i\hbar\partial_t\psi=\left[-\frac{\hbar^2}{2m}\Delta
+\frac{m\omega_\perp^2x_\perp^2+m\omega_z^2z^2}{2}
+i\hbar\frac{\omega_c}{2}(x\partial_y-y\partial_x)\right]\psi
\end{align}
\end{widetext}
The solutions of this equation were given, for example in \cite{cgv}, but for our purpose it is more convenient to use a different notation.

Stationary solutions of the Schr\"odinger equation will be constructed by the separation of variables. The $xy$ part is essentially that of a particle in the uniform magnetic field. These solutions are described in terms of the associated Laguerre polynomials (cf., for example, \cite{bal,bck,kmmr}). The $z$ part corresponds to the one-dimensional harmonic oscillator whose solutions are given in terms of the Hermite polynomials. Thus, the energy eigenfunctions in the Penning trap, are labeled by three quantum numbers $n,l$ and $n_z$,
\begin{widetext}
\begin{align}\label{eigf}
\psi_{nln_z}\!\!=\sqrt{N_{nln_z}}\!
\exp\left[-i\frac{E_{nln_z}t}{\hbar}\right]\!
\exp\left[-\frac{m\omega_\perp x_\perp^2+m\omega_z z^2}{2\hbar}\right]
(x+iy)^l\,L_n^l\! \!\left[\frac{m\omega_\perp x_\perp^2}{\hbar}\right]\!
H_{n_z}\!\left[\sqrt{\frac{m\omega_z}{\hbar}}z\right],
\end{align}
\end{widetext}
where the normalization factor and the energy eigenvalues are,
\begin{align}\label{eigv}
N_{nln_z}&=\frac{n!(m/\hbar)^{l+3/2}\omega_\perp^{l+1}
\omega_z^{1/2}}{2^n\pi^{3/2}(n+l)!n_z!},\\
E_{nln_z} &=\hbar\left[\omega_\perp(2n+l+1)-l\frac{\omega_c}{2}+(n_z+1/2)\omega_z\right].
\end{align}
The probability density for these solutions does not depend on time and the probability current for $l>0$ is flowing in the $xy$ plane in stationary circles around the origin. There seem to be no connection between $\psi_{nln_z}$ and the intricate classical trajectories depicted in Fig.~1. In the next Section we establish a relation between the wave functions and classical trajectories.

\section{Classical vs. quantum description}

There is an essential difference between the description in terms of classical trajectories and the description in terms of the wave functions. In both cases the superposition principle is valid; the superposition of two solutions of the equation of motion is also a solution. However, in the quantum theory it acquires a new meaning.

Let us consider two classical trajectories that differ only in the overall sign. The superposition of two such trajectories produces a trivial result: the particle sits motionless at the center. The superposition of two wave function, however, produces a new wave function with properties quite different from the properties of its two components.

Consider the ground state wave function with the injected trajectory,
\begin{widetext}
\begin{align}\label{g}
\psi_{\{\bm{r}(t),\bm{p}(t)\}}(\bm{r},t)=\exp\!\left[-i\frac{E_{000}t
+\frac{1}{2}\bm{r}(t)\!\cdot\!\bm{p}(t)
-\bm{r}\!\cdot\!\bm{p}(t)}{\hbar}\right]
\exp\!\left[-m\frac{\omega_\perp(x-x(t))^2+\omega_\perp(y-y(t))^2+
\omega_z(z-z(t))^2}{2\hbar}\right].
\end{align}
\end{widetext}
Next, we take the superposition (with the phase difference) of two wave functions with opposite trajectories,
\begin{align}\label{sup}
\psi_S(\bm{r},t)=\psi_{\{\bm{r}(t),\bm{p}(t)\}}(\bm{r},t)+i\psi_{\{-\bm{r}(t),-\bm{p}(t)\}}(\bm{r},t).
\end{align}
Even for the simplest choice of the trajectories this solution reveals the difference between classical and quantum behavior. Choosing in (\ref{sol}) only two initial values, $p_{x0}=p$ and $p_{z0}=q$, different from zero, we obtain,
\begin{subequations}\label{avt0}
\begin{align}
x(t)&=\lambda_\perp\cos\left(\frac{\omega_c}{2}t\right)
\sin(\omega_\perp t),\\
y(t)&=-\lambda_\perp\sin\left(\frac{\omega_c}{2}t\right)
\sin(\omega_\perp t),\\
z(t)&=\lambda_z\sin(\omega_z t),
\end{align}
\end{subequations}
where $\lambda_\perp=p/(m\omega_\perp)$ and $\lambda_z=q/(m\omega_z)$.
The center of the wave packet for the superposition (\ref{sup}) moves according to the equations,
\begin{subequations}\label{avt}
\begin{align}
x_S(t)&=C\lambda_\perp
\cos\left(\frac{\omega_c}{2}t\right)\cos(\omega_\perp t),\\
y_S(t)&=-C\lambda_\perp\sin\left(\frac{\omega_c}{2}t\right)
\cos(\omega_\perp t),\\
z_S(t)&=C\lambda_z\cos(\omega_z t).
\end{align}
\end{subequations}
This is quite similar to the classical trajectory (\ref{avt0}), but the difference in the shape of the wave packet $\psi_S(\bm{r},t)$ is huge. Not only the central moments (\ref{mom}) oscillate but there is a quantum entanglement between the oscillations in the $xy$ and $z$ directions,
\begin{widetext}
\begin{subequations}
\begin{align}\label{cm}
\langle(x-x(t))^2\rangle&=\frac{\hbar}{2m\omega_\perp}+
\lambda_\perp^2\cos^2\left(\frac{\omega_c}{2}t\right)\left(\sin^2(\omega_\perp t)-\exp\left(-2m(\lambda_\perp^2\omega_\perp+
\lambda_z^2\omega_z)/\hbar\right)\cos^2(\omega_\perp t)\right),\\
\langle(y-y(t))^2\rangle&=\frac{\hbar}{2m\omega_\perp}+
\lambda_\perp^2\sin^2\left(\frac{\omega_c}{2}t\right)\left(\sin^2(\omega_\perp t)-\exp\left(-2m(\lambda_\perp^2\omega_\perp+
\lambda_z^2\omega_z)/\hbar\right)\cos^2(\omega_\perp t)\right),\\
\langle(z-z(t))^2\rangle&=\frac{\hbar}{2m\omega_z}+
\lambda_z^2\left(\sin^2(\omega_z t)-\exp\left(-2m(\lambda_\perp^2\omega_\perp+
\lambda_z^2\omega_z)/\hbar\right)\cos^2(\omega_z t)\right).
\end{align}
\end{subequations}
\end{widetext}
The constant terms in these formulas are the values of the central moments for the wave functions $\psi_{\{\bm{r}(t),\bm{p}(t)\}}$.

We can also use the wave functions $\psi_{\{\bm{r}(t),\bm{p}(t)\}}$ to define the distance between trajectories. In the classical theory there is no obvious definition of the distance between two trajectories, but in quantum theory the distance between two states is well defined in terms of fidelity, $f=|\langle\psi_1|\psi_2\rangle|^2$. By injecting two trajectories into the wave function of the ground state we obtain two wave functions whose fidelity is,
\begin{align}\label{f}
f=\exp\!\left[-\frac{Q(\bm{p}_1(t)-\bm{p}_2(t),
\bm{r}_1(t)-\bm{r}_2(t))}{\hbar}\right],
\end{align}
where the quadratic form $Q$ is,
\begin{align}\label{Q}
Q\left(\bm{p}(t),\bm{r}(t)\right)=\frac{p_\perp^2}
{2m\omega_\perp}+\frac{p_z^2}{2m\omega_z}+
\frac{m\omega_\perp x_\perp^2}
{2}+\frac{m\omega_zz^2}{2}.
\end{align}
We omitted the time dependence in this formula because $Q$ is a constant of motion; its time independence is inherited from the time independence of fidelity. Since the quadratic form $Q$ is positive definite, it may be used to define the distance $d(tr_1,tr_2)$ between two trajectories $tr_1$ and $tr_2$,
\begin{align}\label{d}
d(tr_1,tr_2)=\sqrt{Q\left(\bm{p}_1-\bm{p}_2,
\bm{r}_1-\bm{r}_2\right)}.
\end{align}
This distance is a purely classical quantity even though is has been derived with the use of quantum mechanical wave functions.

\section{Conclusions}

The results presented in this work serve a dual purpose. On one hand they add new analytic solutions of the Schr\"odinger equations that have a clear physical relevance. These solutions bridge the gap which existed up to now between the classical and quantum descriptions of the motion of particles in the Penning trap. On the other hand they give in this case a precise meaning to the notion of the classical-quantum correspondence. They may also be viewed as an explicit example of the exact validity of the Ehrenfest theorem.

\end{document}